# Novel Methods to Create Multielectron Bubbles in Superfluid Helium

J. Fang[1], Anatoly.E. Dementyev[1], J. Tempere[1,2], and Isaac F. Silvera[1]

[1]Lyman Laboratory of Physics, Harvard University, Cambridge MA 02138

[2]TFVS Universiteit Antwerpen, Groenenborgerlaan 171, 2020 Antwerpen, Belgium

An equilibrium multielectron bubble in liquid helium is a fascinating object with a spherical two-dimensional electron gas on its surface. We describe two ways of creating them. MEBs have been observed in the dome of a cylindrical cell with an unexpectedly short lifetime; we show analytically why these MEBs can discharge by tunneling. Using a novel method, MEBs have been extracted from a vapor sheath around a hot filament in superfluid helium by applying electric fields up to 15 kV/cm, and photographed with high-speed video. Charges as high as $1.6 \times 10^{-9}$ C ($\sim 10^{10}$ electrons) have been measured. The latter method provides a means of capture in an electromagnetic trap to allow the study of the extensive exciting properties of these elusive objects.

## I. Introduction

Multielectron bubbles (MEBs), which can range in diameter from nanometers to millimeters, were first observed over 30 years ago by Volodin, Khaikin, and Edel'man [1] using high-speed photography. Although they possess a plethora of exciting predicted phenomena including superconductivity, vortex states, sonoluminescence, magnetic properties such as the quantum Hall effect, electron localization, instabilities, etc., recently reviewed by Tempere,

Silvera, and Devreese [2], there have been no studies of these phenomena, as the lifetimes of the observed MEBs were short, of order a few ms. In this paper we report our recent advances using novel techniques to produce MEBs in configurations where the lifetimes can be greatly extended.

In the original method [1] electrons were deposited on a superfluid liquid helium surface to form a 2D electron gas, bound to the surface by the weak attractive image charge in the helium. In order to increase the surface charge density, a flat electrode with a positive electric potential was placed under the helium surface. When the electric field exceeded ~3-4 kV/cm, the helium surface became unstable and subsumed of order $N = 10^7$-$10^8$ electrons in the form of MEBs that rapidly moved to the electrode where they were annihilated. Later, Albrecht and Leiderer [3] studied MEBs in normal helium by photographic techniques, at temperatures above the helium lambda point, and measured their velocities to be ~10 cm/s.

The radius of an equilibrium MEB, called the Coulomb radius varies, depending on the charge $Ne$. Shikin [4] determined this radius by minimizing the energy, including the surface tension of helium and the Coulomb repulsion of the electrons, with a term for quantum mechanical localization. This yielded the Coulomb radius $R_C = \left( e^2 / 16\pi\varepsilon_{He}\sigma \right)^{1/3} N^{2/3}$, where $\sigma$ is the surface tension of helium and $\varepsilon_{He}$ its dielectric constant. For N=$10^4$ electrons the Coulomb radius is 1.064 microns. Shikin pointed out that at the Coulomb radius MEBs should be unstable to deformation and fissioning.

Waves on a flat surface of helium are called ripplons; on a spherical surface, spherical ripplons. The various angular modes of oscillation have deformations described by spherical harmonics, $Y_{\ell m}(\theta,\phi)$ and are indexed by $\ell, m$. At zero pressure the mode $\ell = 2$ has zero restoring force and is unstable, based on a harmonic approximation. Tempere, Silvera, and Devreese [5] investigated the effect of pressure on the statics and dynamical ripplon modes and found that a negative pressure stabilized the bubble (the $\ell = 2$ mode frequency becomes non-zero) and higher order mode frequencies increase rapidly with pressure, then turn over and go to zero frequency.  Salomaa and Williams [6] used density functional theory to show that the width of the electron gas on the bubble surface is ~1 nm, and that non-linearities might stabilize the MEB against the $\ell = 2$ mode instability at zero pressure.  Fissioning of MEBs was studied [7] using Bohr's liquid drop model, and when charge redistribution is taken into account [8] it is seen that positive pressure is detrimental to bubble stability.  Thus, it is important to create and understand the stability of MEBs by producing them in long-lived states so that their remarkable properties may be accessible to experimental study

## I.  Experimental Production of MEBs

### i.  Sessile Bubbles

We have produced MEBs by two methods, with the objective of stabilizing them for long periods of time.  In the first method [9] a cylindrical brass cell with a dome shaped top, shown in Fig. 1, has a reservoir connected to the cell by a flexible capillary tube so that the helium level in the cell can be raised and lowered. With the helium at temperatures of order 1.3 K and the level down in the cell, all surfaces above the bulk helium are covered by a superfluid film that should be several hundred angstroms thick, depending on the vertical distance to the surface of the bulk helium.  There is a 1 eV barrier for electrons to enter the bulk helium [10] so that if electrons (produced with a source, either thermionic emission [11] or field emission tips [12]) are introduced into the vapor region of the cell, they should cover the helium surfaces and be confined to the vapor phase.  As the bulk helium surface is raised the electrons are corralled into a smaller and smaller volume until the helium level reaches the top of the dome where a minimum energy surface, an MEB, should be formed and localized by buoyancy; the pressure can be varied by changing the He level.

In the original cell, the top of the dome was a concave glass lens with a transparent metallic coating (to be used for measuring the induced image charge, and thus the charge of the MEB). An MEB could be illuminated with light from an optical fiber and imaged with a low temperature microscope on a coherent fiber optic bundle (30,000 fibers with 3.5 micron diam.), so that the image could be transferred outside of the cryostat for viewing. However, MEBs could not be detected either in this configuration or later with an uncoated lens. Our low temperature electron collector [13] showed a strong production of electrons in the cell. Measurements elsewhere [14, 15] have shown that although the helium film under electrons on metallic surfaces is thinned due to attraction of electrons by image charges in the metal, the helium surface maintains substantial thickness, ~50-100 Å, and the electron surface density only slowly decays, so that we should be able to load electrons into the vapor phase and produce MEBs. We suspected severe loss of electrons with the thermionic source turned on due to heating that can burn the helium film off of surfaces, allowing electrons to flow directly into the metallic walls of the cell. To overcome the problem of "film burning" we installed a copper "holding plate" in the bottom of the cell with a positive potential. This was submerged a few millimeters under the bulk helium when the source was operating, so that ~ $10^9$-$10^{10}$ electrons could be collected above this plate when an intense source of electrons was activated. With the source off and surfaces covered with a film, the electrons could be released and the helium level could be raised.

Using this technique we succeeded in producing large MEBs, but with a surprise concerning localization of the bubble. At the beginning of a run (repetitive cycles of loading of electrons and raising the helium level) there were no results, but then large MEBs with diameters of order 1-1.5 mm ( $N \sim 1-2 \times 10^8$ electrons) were visually observed with the microscope. However, these did not localize in the top of the dome as expected, but moved across the field of view of the fiber optic bundle in times of order 1 s. This corresponds to lifetimes two to three orders of magnitude longer than in earlier observations, when MEBs were formed by the instability of flat surfaces.

The pressure head of the helium was approximately zero, indicated by our built in helium level detector (concentric stainless steel tubes that form a capacitor to measure the helium level, not shown in Fig. 1). Two electrodes designed to produce electric fields to steer MEBs were mounted in the cell in the dome region and these caused the bubbles to change their direction of motion, confirming that they contained charge, but not localizing them. After some cycles there was deterioration and MEBs could no longer be observed. The cell was warmed to ~20 K, and then cooled, and again MEBs could be produced after some loading cycles, but could not be localized in the dome. This observation was briefly reported earlier [2], but not understood. The thickness of the helium film above the MEB was calculated to be ~30-70 Å, depending on N, too thick to allow electron loss via tunneling. Below we describe improvements in the analysis that have a dramatic effect on the film thinning and explain the observations. The most important effect is to consider the interaction of an electron with all of the image charges, not just its own.

Electrons in an MEB "see" an image charge in the surface of the dome, producing a pressure on the surface of the helium film that thins the film. The thickness $d$ of the helium film between an MEB and the surface above it is determined by force balance. Buoyancy and the electrostatic attraction to the image charges draw the bubble towards the surface, whereas the Van der Waals interaction favors pushing the bubble away from the surface. The energy of an MEB with radius $R_N$, shown in Fig. 2, containing $N$ electrons, and whose center is situated at a distance $D \equiv \delta \times R_N$ from the cell surface above it, is given by

$$E = \pi \alpha \left[ \frac{2\delta}{\delta^2 - 1} + \log\left(\frac{1-\delta}{1+\delta}\right) \right] + \frac{4\pi}{3} \rho g R_N^4 \delta - \frac{(Ne)^2}{4\pi\varepsilon_{He} R_N} \frac{1}{2\delta} \tag{1}$$

Here, the first term represents the Van der Waals energy, with α the Van der Waals constant defined such that $\alpha dV/z^3$ is the energy of an infinitesimal volume $dV$ of helium at a distance $z$ from the surface. The second term represents buoyancy and the last term represents the electrostatic attraction of the bubble to the metal surface (ρ is the helium density, and g=9.81 m/s$^2$). For dielectric substrates, this last term should be multiplied by (1−ε)/(1+ε) where ε is the relative dielectric constant of the substrate. Charge redistribution of electrons along the surface of the MEB add higher-order multipole terms to the electrostatic interaction, but the effect of these terms is limited to a few percent: the redistribution of charge costs direct Coulomb energy in the MEB itself.

For all except the largest bubbles (with $N \geq 10^8$ electrons) the buoyancy energy can be neglected in Eq. 1. The helium film thinning due to MEBs, obtained from minimizing the above energy with respect to δ, is found to be much more severe than for a 2D electron gas on a flat surface [14]. In the approximation that the film thickness $d = D - R_N$ (typically on the order of a nm) is much smaller than the bubble radius, the energy minimization can be done analytically. We find

$$d = \frac{1}{4} \sqrt{\frac{\alpha}{2\sigma}} \tag{2}$$

where σ is the surface tension of the helium that appears through the expression for the bubble radius, $R_N$ as a function of $N$. For a dielectric rather than a metal, this result should be multiplied by $\sqrt{(1+\varepsilon)/(1-\varepsilon)}$. The film thickness is representative for bubbles up to ~$10^8$ electrons (larger bubbles will have an even smaller value of $d$), and independent of the number of electrons. As the number of electrons is increased, the growth of the electrostatic interaction between charges and image charges is cancelled by the growth in bubble radius separating the charges. For values $\alpha$=8.12x10$^{-22}$ J [16] for copper, the film thickness is reduced to 5.0 Å, which is thin enough to allow electrons to tunnel out of the MEB into the surface. Also for glass with $\varepsilon$=2.25 and $\alpha$=3.92x10$^{-22}$ J [17, 18] the resulting thickness of 5.6 Å is too small to prevent electrons from tunneling into the glass surface, where they are trapped and localized. As this surface charge builds up, the resulting large electric fields at the surface repel MEBs, overcoming the buoyant forces. Presumably the trapped charge continues to build up during "firing" of the electron source and eventually MEBs are repelled from the field of view. Warming of the cell to 20 K allows the trapped electrons to escape and the process can start anew. Details of these calculations will be presented elsewhere.

Although this technique has greatly extended the lifetime of MEBs, having been observed a few hundred times longer than in the original method, it is insufficient for studying the properties of MEBs under static equilibrium conditions. One might be able to improve such a system and localize the MEBs with carefully designed electrodes, but the continually changing conditions create challenges, and another technique for producing MEBs isolated from non-helium surfaces was developed.

### ii. Electric Field Extraction of Bubbles

It was shown by Spangler and Hereford [19] that if a current is passed through a tungsten filament submerged in liquid helium, the filament can glow at temperatures of a few thousand K. The filament is insulated from the liquid by a vapor sheath that is formed around it, the sheath also containing electrons due to thermionic emission. They were able to draw electrons out of the sheath to an anode in the form of single electron bubbles. This phenomenon of formation of a vapor sheath with i-V (current-voltage) characteristics completely different from that of a filament in vacuum has recently been studied by Silvera and Tempere [11] who described the transition to the state with a sheath as a first-order phase transition. For a straight wire the sheath forms a small diameter cylinder around the wire, tightly bonded to the wire. However, this can be expanded in volume by modifying the geometry or shape of the glowing filament.

We have made a filament in the shape of a loop operating at ~1.5 K in superfluid helium, shown in Fig. 3, to form a sheath in the shape of a large bubble filled with electrons. The filament was placed in a helium-filled glass cryostat with a liquid nitrogen jacket. The back illuminated filament could be observed under the modest magnification of a long working-distance microscope, through the four strip windows of the glass cryostat. Initial efforts to break an electron filled bubble loose from the sheath were futile as the sheath remained tightly tethered to the filament when helium was below the lambda point and would only tend to radially oscillate as the current or the temperature of the filament was increased. At the lambda point the helium can no longer efficiently carry away the heat generated by the filament and torrents of bubbles are formed and rise up in the liquid helium column. We measured the charge contained in bubbles annihilated by a copper collector at the top of the helium column to be ~$10^7$ electrons [13]. In this configuration we were unable to produce isolated electron filled bubbles, even when pulsing the filament; most of the bubbles were produced at warmer, but non-glowing parts of the filament, so they would contain few electrons. Furthermore, our objective was to produce isolated MEBs in superfluid helium. We decided to extract MEBs from the sheath using forces generated by strong electric fields.

A planar anode was placed a few mm to the side of the filament. Clearly, fields larger than ~ 3 kV/cm are needed for extraction from the curved surface of the sheath, as this is the field value for which a flat surface becomes unstable. The anode was biased with a positive voltage so that dc electric fields as high as 15 kV/cm could be produced, either continuously or pulsed on and off. Using this approach we have succeeded in producing MEBs that we have photographed. The creation of MEBs was captured using high-speed photography, and is shown in Fig. 4. A bubble is pulled out of the sheath and moves to the right in the direction of the electric force on the electrons in the bubble, an MEB. Two forces act on the sheath or bubbles: the electric force and the buoyant force. By varying the electric field, MEBs can be created that move towards the electrode, but continue to rise up in the helium. If the electric field is too strong, then the sheath can stretch and extend to the anode, as shown in Fig. 5. When the sheath connects the loop and anode, current flows and there is large dissipation causing a mini explosion in the helium, which settles back down in a few seconds.

The creation of the bubble shown in Fig. 4 is quite violent and it oscillates in size becoming smaller, larger, smaller, and then too small to see (not shown). In principle one can determine the charge in the bubble from the oscillation period, using the Rayleigh-Plesset equation for radial dynamics of a bubble [20], modified to take into account the electrostatic effect of the charges in the bubble:

$$R\ddot{R} + \frac{3}{2}\dot{R} = \frac{1}{\rho}\left[\frac{2\sigma}{R} + p - \frac{(Ne)^2}{8\pi\varepsilon_{He}R^4}\right].$$

(4)

This can be integrated to obtain

$$\dot{R}^2 = -\frac{1}{2\pi\rho r^3}\left\{U[R_b(t)] - U[R_b(0)]\right\},$$

(5)

with

$$U(R) = \frac{(Ne)^2}{2\varepsilon_{He}R} + 4\pi R^2 \sigma + \frac{4\pi}{3} R^3 p.$$

(6)

Eq. (5) gives rise to a periodic behavior of $R(t)$, where the minimum and maximum radius are the turning points of the potential $U$. We observe one growth-collapse cycle, and during this cycle the maximal bubble radius is $R_{max}$=0.5 mm and the minimal radius is $R_{min}$=0.2 mm. We use an average radius since at this stage the bubble is not spherical. Since these radii are the turning points of the potential (5), we can estimate the number of electrons in the bubble through solving for $N$ in $U(R_{max}) = U(R_{min})$, and obtain $N$=1.4×10$^8$. Whereas the maximum average radius could be determined fairly accurately, the minimum radius is an estimate: due to the fast dynamics near the minimum, the true minimal radius could have been reached in between two frames. This means that our estimated charge is an upper bound for $N$.

In Fig. 6 we show the comparison of the experimental radii to two theoretical curves obtained from Eq. (4). The dashed curve is for N=10$^7$ electrons and the full curve is for N=10$^8$ electrons. From this it is clear that the determination of the period of oscillation had insufficient precision to refine the estimate. Moreover, successive growth and collapse cycles reduce the radius of the bubble and this points to a possible loss of electrons each time the bubble reaches minimum radius. Creation of bubbles in this manner was most successful when the pressure due to the helium head was quite small; in this case the head was several millimeters. In a separate experiment MEBs were extracted and allowed to rise to a collector. Current pulses were measured with an instrumentally determined width of ~1 ms and charges up to 10$^{10}$ electrons were found, shown in Fig. 7.

Thus, we have succeeded in producing MEBs in two ways, with the dome shaped cell and the filament. In first case the pressure on the bubble was around zero. The bubbles were observed for ~1 s as they moved through the field of view, but the lifetime could be much longer and may point to stability when the pressure head is around zero or lower. The electric field extraction method seems to be the most attractive for further localizing MEBs. It is our intention to develop an electromagnetic trap above the filament and trap MEBs. This will be done in a cell with a greatly improved optical path to enhance resolution and viewing, as well as having the capability of controlling the helium pressure, including producing negative pressures in the cell. With MEBs localized in a trap, one can then study their stability with respect to pressure, tunneling by introducing a macroscopic surface into the trap, and many of the exciting properties that have been predicted for MEBs.

We thank Eugene Gordon, Jozef Devreese, and Shanti Deemyad for useful discussions and some assistance. This research was supported by the U.S. Department of Energy, Grant DE-FG02-ER45978 and by the NOI BOF UA 2004, the FWO-V projects Nos. G.0435.03, G.0274.01, G.0356.06, G.0115.06, G.0180.09N, G.0370.09N.

**Figure Captions**

Fig. 1.  A dome-topped cell for trapping MEBs in superfluid helium.

Fig. 2.  A sessile MEB with radius R immersed in liquid helium with its center a distance D from a metallic or dielectric surface. The charges (+ and -) are not uniform as the energy can be lowered by charge redistribution. For small bubbles, deformation from sphericity can be ignored.

Fig. 3.  A vapor sheath filled with electrons around a glowing tungsten filament in superfluid liquid helium. The tungsten loop diameter is ~1 mm.

Fig. 4  Selected frames from a video (6400 frames/s or 0.15 ms/frame) of an MEB extracted from the sheath around a tungsten loop with diameter ~1 mm.  An electrode 3 mm to the right (not shown) creates an electric field up to 15 kV/cm with respect to the loop.  An MEB is seen oscillating in size as it rises. The filament is back illuminated. The dark "clouds" (out of focus) are slowly rising vapor bubbles in the liquid nitrogen bath, surrounding the helium bath.

Fig. 5.  Same as Fig. 4, except the electric field is higher. A bubble being extracted extends continuously from the filament to the positively biased anode. When contact is made (frame at 4.06 ms) a large current flows and a spectacular explosion occurs in the superfluid, but with no damage.

Fig. 6.  The bubble radius, as a function of time. The experimentally observed radii for the different timeframes are indicated by points with error bars stemming from the fact that the bubbles are not precisely spherical. The curves show the result of the Rayleigh-

Plesset equation extended to charged bubbles, for a charge of N=$10^7$ electrons (dashed) and N=$10^8$ electrons (full line).

Fig. 7.  Charge collected from an MEB rising through the superfluid helium (T=1.5 K) to a collector submerged in the helium.  The integrated area yields a charge of ~$1 \times 10^{10}$ Coulombs.  For this bubble the amplifier was slightly saturated at the peak.

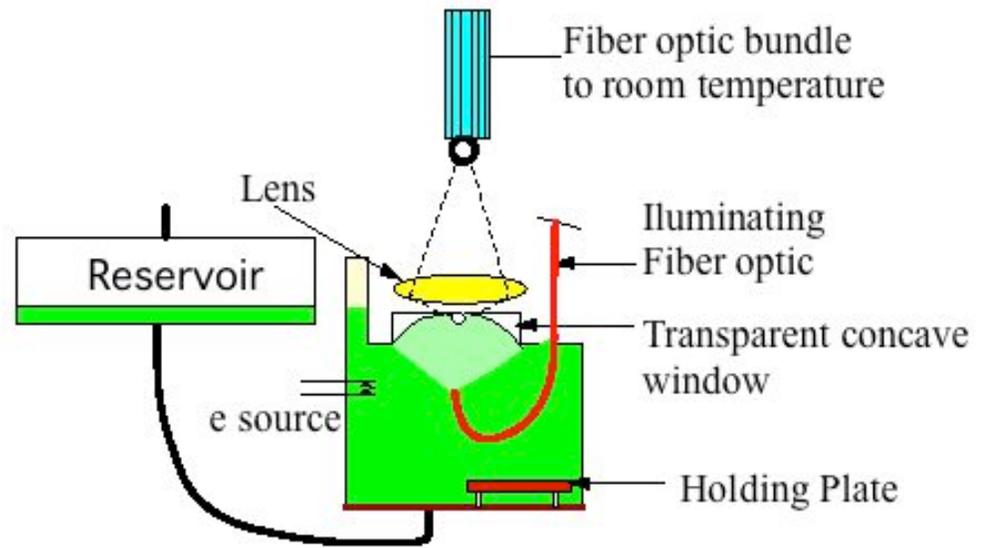

Fig. 1.

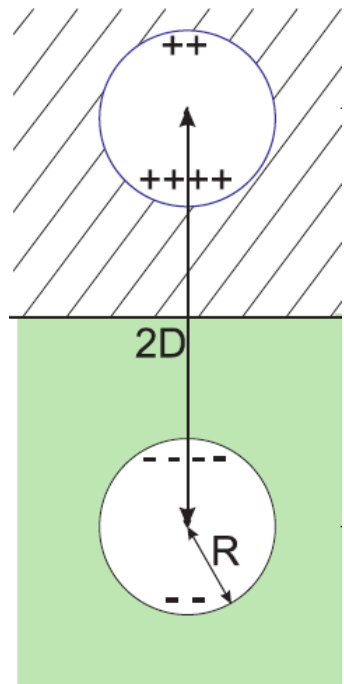

Fig.2.

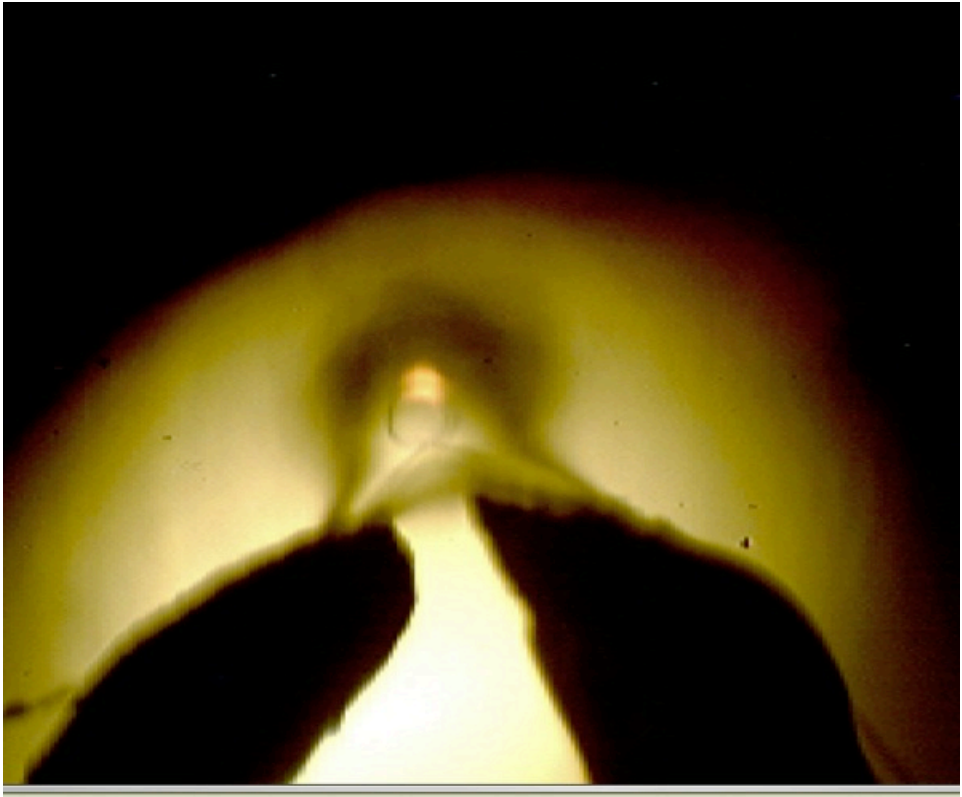

Fig. 3

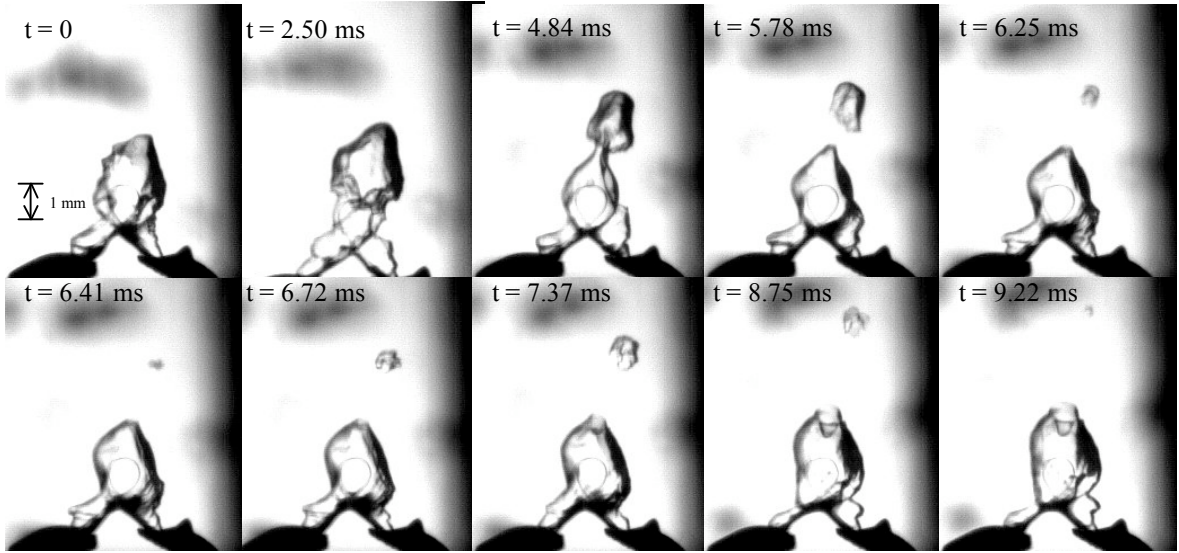

Fig. 4

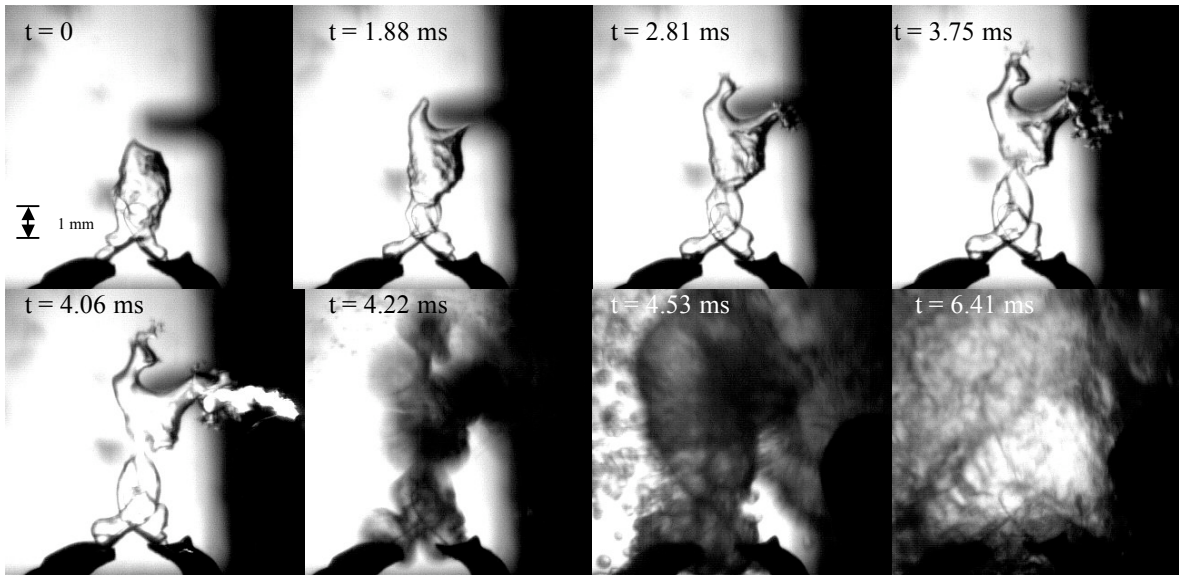

Fig. 5.

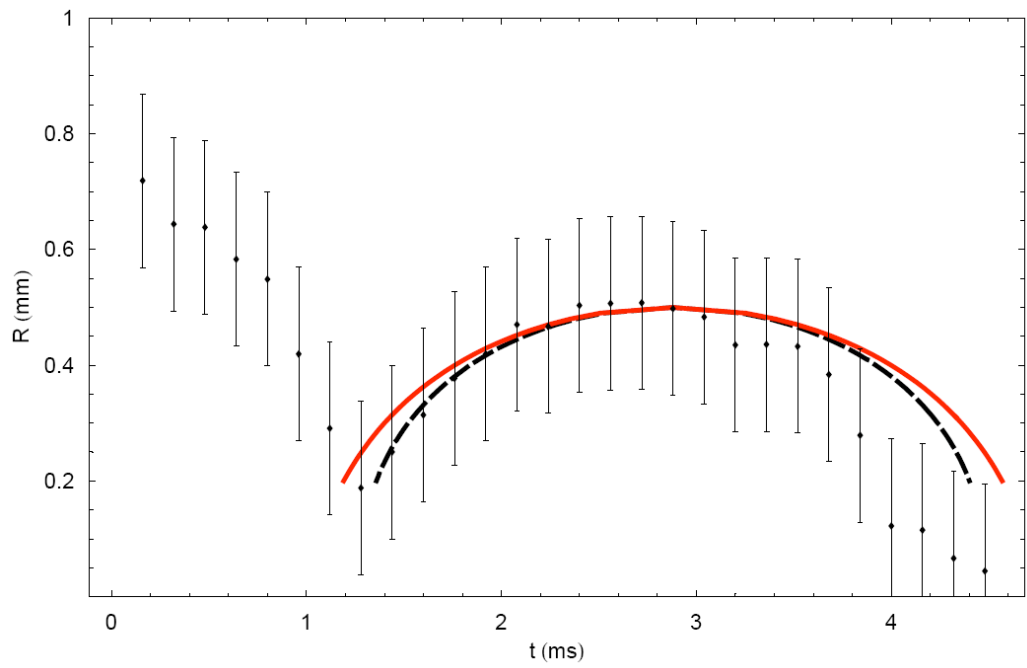

Fig. 6

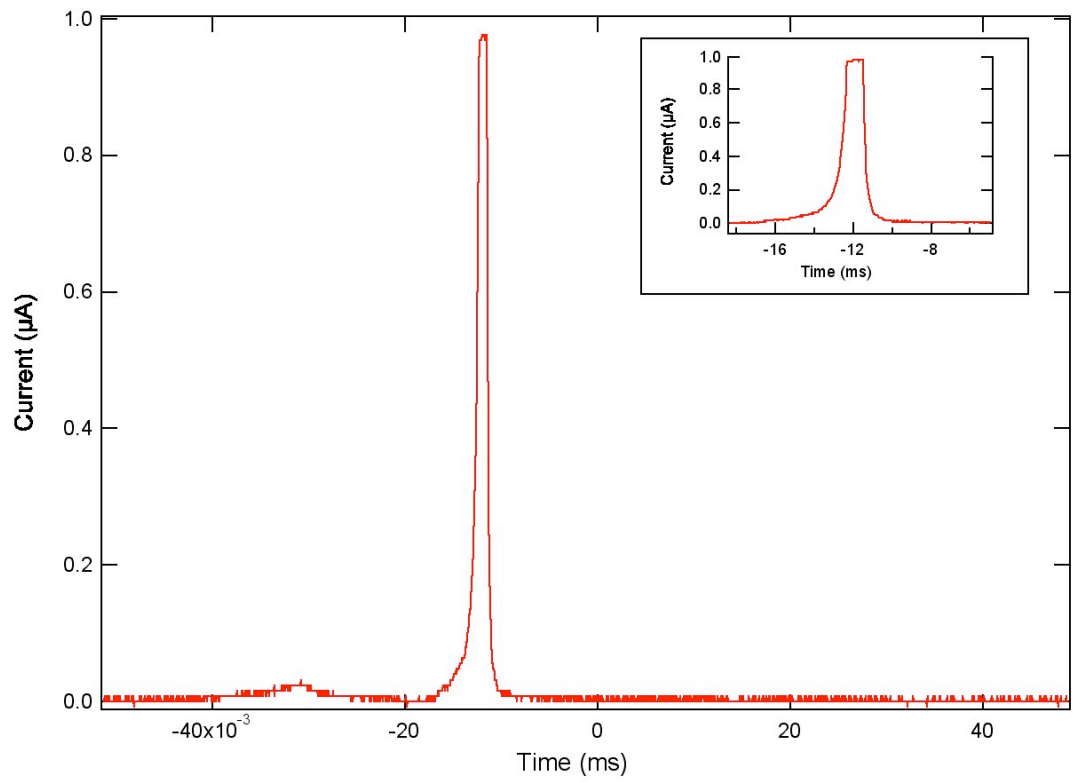

Fig. 7.